\documentclass[12pt]{article}
\usepackage{amsmath}
\usepackage{amssymb}
\usepackage{amsfonts}
\newcommand{\nc}{\newcommand*}

\nc{\beq}{\begin{equation*}}
\nc{\eeq}{\end{equation*}}
\nc{\beqa}{\begin{eqnarray*}}
\nc{\eeqa}{\end{eqnarray*}}
\nc{\lcr}[2]{{\left[{#1},{#2}\right]}}    
\nc{\ket}[1]{{\vert{#1}\rangle}}          
\nc{\ccr}[2]{\left[{#1}\,,\,{#2}\right]}  
\nc{\der}[1]{\frac{\rm d}{{\rm d}{#1}}}   
\nc{\Szi}[1]{\sum_{ #1 = 0 }^{\infty}}
\def\da{a^{\dagger}}

\def\half{{\frac{1}{2}}}
\def\H{\mathcal{H}}
\def\cN{\mathcal{N}}
\def\cA{\mathcal{A}}
\begin{document}
$${}$$
\begin{flushright}
{Translation from Russian}\\
{of the talk given on the }\\
{Chebyshev Workshop, Obninsk, 2002}
\end{flushright}
$${}$$
\centerline{\huge\bf Coherent states}
\medskip

\centerline{\Large\bf  and}
\medskip

\centerline{{\huge\bf Chebyshev polynomials}
\large\footnote{This research is
supported in part by RFFI grant no. 00-01-00500.}}

\bigskip
{\flushleft{\large\bf V.V.Borzov,${}^*$  E.V.Damaskinsky${}^{**}$}}
\smallskip
{\flushleft{${}^*$ St.-Petersburg University of Telecommunications\\
E-mail: vadim@VB6384.spb.edu}}
\vspace{-.1cm}
{\flushleft{${}^{**}$St.-Petersburg University of the Defence Engineering
Constructions\\
E-mail:  evd@pdmi.ras.ru}}
\bigskip

The important role of coherent states \cite{1,2} in the contemporary
quantum physics (especially, in quantum optics) is well-known.
In particular, coherent states are essentially used the
definition and evaluation of functional integrals~\cite{3}.
The concept of coherent states has appeared, useful also in,
the mathematical physics, in the theory of representations and in some
other branches of mathematics.

Introduced for boson oscillator (the Heisenberg group)
the coherent states now are defined for
wide class of quantum physical systems (including
quantum fields), and also for systems, connected to other groups
(including the supergroups). The coherent states can be defined
for quantum groups, and also for various deformations (generalizations)
of the exponents. The extensive and rather full bibliography
of the theoretical researches concerning with coherent states,
squeezed states and their numerous generalizations, is contained
in~\cite{5}.

There are several definitions of coherent states
\begin{enumerate}
\vspace{-8pt}\item as eigenstates of annihilation operator:
$a\ket{z}=z\ket{z},\quad z\in{\mathbb C}$\,
({\it Barut - Girardello coherent states});
\vspace{-10pt}\item as outcome of action of unitary displacement operator
$D(z)=e^{z\da-z^*a}$ on the fixed vector of the Hilbert space
(usually the Fock vacuum  $\ket{0}$):
$\ket{z}=D(z)\ket{0}$ ({\it Perelomov coherent states});
\vspace{-10pt}\item as the states, minimizing the Heisenberg (or
Schr\"odinger - Robertson)
uncertainty relations;
\vspace{-10pt}\item as the states, satisfying to small number of the
natural conditions~\cite{4} ({\it Klauder-Gazeau coherent states}).
\end{enumerate}

In the case of the boson oscillator all these definitions generates
the same set of coherent states, however it is not so in more
general cases.

In the works of the authors~\cite{6,6a,7,8} the new approach to a definition of
coherent states is suggested. This approach is connected with the
given in the work~\cite{9} construction of generalized oscillator algebras,
connected with arbitrary system of the
orthonormalized polynomials, which play in this case the same role,
as Hermite polynomials in the case of standard boson oscillator.
Namelly, for the given system of orthogonal
polynomials by a canonical way one can define the oscillator-like
system and introduce the position, momentum, Hamiltonian operators
(which spectrum is determined by the coefficients
of recurrent relations  of these orthogonal polynomials).
The related ladder (creation and annihilation) operators, satisfy the
commutation relations of the deformed boson oscillator algebra. This
allows in the case of classical orthogonal polynomials (as well as for
their various $q$-analogs) to construct the coherent states of such
oscillator-like systems.

The suggested method for definition of coherent states is used
in~\cite{6,7,8} for construction of Barut - Girardello and Klauder - Gazeau
coherent states connected with  the Hermite, Laguerre and Legendre
polynomials. The case of Perelomov-type coherent states, which is more
difficult technically, is in a stage of completion and will be presented in
other publication.

In the present note the coherent states of the Barut - Girardello type
are defined for the oscillator-like systems, connected with the Chebyshev
polynomials $T_n(x)$ and $U_n(x)$ of the 1-st and 2-nd kind. It is natural
to call such systems as Chebyshev oscillators.

Let us recall that Chebyshev polynomials  $T_n(x)$ of the 1-st kind
 are defined for $x\in [-1;1]$ by the relation
$$
T_n(x)=\cos(n\theta)=\cos(n\arccos x),\quad n=0,1,\ldots,\qquad
\theta=\arccos x .
$$
These polynomials fulfill the recurrent relations
$$
T_{n+1}(x)=2xT_{n}(x)-T_{n-1}(x).
$$
The polynomials $T_n(x)$ are orthogonal on a segment $[-1;1]$
with the weight function
$$
 h(x)=\frac{1}{\sqrt{1-x^2}}=\frac{1}{\sin\theta},\qquad
x\in (-1;1) .
$$

The Chebyshev polynomials $U_n(x)$ of the 2-nd kind are defined
for $x\in [-1;1]$  by the relation
$$
U_n(x)=\frac{\sin((n+1)\arccos x)}{\sqrt{1-x^2}}=
\frac{\sin((n+1)\theta)}{\sin(\theta)},\, n=0,1,\ldots,\quad
\theta=\arccos x .
$$
These polynomials satisfy the same recurrent relations
$$
U_{n+1}(x)=2xU_{n}(x)-U_{n-1}(x).
$$
and are orthogonal on a segment $[-1;1]$ with weight function
$$
 h(x)={\sqrt{1-x^2}}={\sin\theta},\qquad x\in [-1;1].
$$

Let's consider the Hilbert spaces
$$
{\H}_1=\text{L}^2\left([-1,1];\frac{{\rm d}x}{\pi\sqrt{1-x^2}}\right)\,,
\qquad
{\H}_2=\text{L}^2\left([-1,1];\frac{2\sqrt{1-x^2}{\rm d}x}{\pi}\right)
$$
which basises are formed by the Chebyshev polynomials of the 1-st
and 2-nd type
\beqa
\Psi^{(1)}_{n}(x)&=&\sqrt{2}T_n(x),\quad n\geq 1;\qquad
\Psi^{(1)}_{0}(x)=T_0(x)=1;\\
\Psi^{(2)}_{n}(x)&=&U_n(x),\quad n\geq 0,
\eeqa
respectively. The recurrent relations ($j=1,2$)
\beq
x\Psi^{(j)}_{n}(x)=b_{n}\Psi^{(j)}_{n+1}(x)+b_{n-1}\Psi^{(j)}_{n-1}(x),
\quad n\geq 0, \quad b_{-1}=0;
\eeq
where
\beq
\Psi^{(j)}_{0}(x)=1;\qquad b_n=\half,\,\,n\geq 1,\,\,
b_0=\frac{1}{\sqrt{2}},
\eeq
defines an action of "coordinate"  operator $X_j$ on the basis elements
of these spaces.

By method circumscribed in~\cite{9} we define the related "momentum"
operators $P_j$ and quadratic Hamiltonians ("energy" operators)
$H_j=X_j^{\,2}+P_j^{\,2}.$

It is not hard to find the eigenvalues of these "energy" operators,
which in both cases have the following form
$\lambda_0=\half,\quad \lambda_n=1,\, n\geq1 .$

Considering Hilbert spaces ${\H}_j$ as Fock spaces of such defined
Chebyshev oscillators we introduce the creation and annihilation operators
\beq
a_j^{\pm}=\frac{1}{\sqrt{2}}\left( X_j\pm iP_j\right)
\eeq
and the operator of a number of the state
$N_j \Psi^{(j)}_{n}(x)=n \Psi^{(j)}_{n}(x).$

Now we denote $B(N_j)$ the operator-valued
functions acting on the basis elements in  ${\H}_j$  as follows
\beq
B(N_j)\Psi^{(j)}_{n}(x)=b_{n-1}^{\,2}\Psi^{(j)}_{n}(x) .
\eeq

It is not hard to check that so defined operators satisfy
the following commutation relations
\beq
\ccr{a_j^-}{a_j^+}=2\left(B(N_j+I_j)-B(N_j) \right),\qquad
\ccr{N_j}{a_j^{\pm}}=\pm a_j^{\pm},
\eeq
and defines the deformed Heisenberg algebra ${\cA}_{\Psi}.$
It is natural to call the Chebyshev oscillator algebra.
We note that Chebyshev polynomials of the 1-st and 2-nd kind
determines the unitary equivalent representations of this algebra
in  Hilbert spaces ${\H}_1$ and ${\H}_2.$

Let's consider a differential operator  $A=(1-x^2)\der{x},$
acting on the basis elements in spaces ${\H}_j$ as follows
\beqa
A\Psi^{(1)}_{n}(x)&=
 &nb_{n-1}\Psi^{(1)}_{n-1}(x)-nb_{n}\Psi^{(1)}_{n+1}(x),\\
A\Psi^{(2)}_{n}(x)&=
 &(n+2)b_{n-1}\Psi^{(2)}_{n-1}(x)-nb_{n}\Psi^{(2)}_{n+1}(x).
\eeqa
Then we have
\beqa
a_j^-\!\!&=&\!\!
 \frac{1}{\sqrt{2}}\left(N_j+I_j\right)^{-1}\left(A+X_jN_j\right)
,\quad j=1,2;\\
a_1^+\!\!&=&\!\!
 \frac{1}{\sqrt{2}}\left(N_1-I_1\right)^{-1}\left(-A+X_1N_1\right);\\
a_2^+\!\!&=&\!\!
 \frac{1}{\sqrt{2}}N_2^{\,-1}\left(-A+X_2N_2+2X_2I_2\right);\\
P_1\!\!&=&\!\!i\left(N_1-I_1\right)^{-1}
 \left(N_1+I_1\right)^{-1}\left( N_1A-X_1N_1\right);\\
P_2\!\!&=&\!\!i\left(N_2\right)^{-1}
 \left(N_2+2I_2\right)^{-1}
\left( N_2A+I_2A+X_2N_2-X_2\right).
\eeqa

The Barut - Girardello coherent states (i.e. the eigenfunctions of
annihilation operator) for the Chebyshev oscillator in the space ${\H}_1$
are defined by a relation
\beq
\ket{z}_1={\cN}^{-1}\Szi{n} z^n\sqrt{2}^{\,n+1}T_n(x),
\eeq
where the normalizing factor is equal
\beq
{\cN}^{2}=\Szi{n}(\sqrt{2}|z|)^{2n}=\frac{1}{1-2|z|^2},\quad
|z|<\frac{1}{\sqrt{2}} .
\eeq

Using a generating function
\beq
\Szi{n}r^{n}T_n(x)=\frac{1-rx}{1-2rx+r^2}, \quad |r|<1,
\eeq
for Chebyshev polynomials of the 1-st kind, one obtains, for $r=\sqrt{2}z,$
\beq
\ket{z}_1=\sqrt{1-2|z|^2}\,
\frac{1-\sqrt{2}zx}{1-2\sqrt{2}zx+2z^2},\quad
|z|<\frac{1}{\sqrt{2}} .
\eeq

The related holomorphic representation in Bargmann type space
consists of functions analytical in a disk $|z|<\frac{1}{\sqrt{2}}.$
The constructed set of coherent states has all standard properties.
The resolution of identity is determined by a measure, representing
$\delta$-function on the boundary of a disk $|z|<\frac{1}{\sqrt{2}}.$

The construction of Klauder-Gazeau coherent states for the Chebyshev
oscillator also don't present any problem. At the same time,
the definition of coherent states of Perelomov type
calls some technical difficulties (connected with a lack
"disentangling formula").
Construction of this type  coherent states for the Chebyshev oscillator
authors hope to finish in a near future.

\end{document}